# Characterizing Tourist Daily Trip Chains Using Mobile Phone Big Data

Yan Luo

silver.luo@connect.polyu.hk

Department of Computing

The Hong Kong Polytechnic University

## ABSTRACT

Tourists tend to visit multiple destinations out of their variety-seeking motivations in their trips. Thus, it is critical to discover travel patterns involving multi-destinations in tourism research. Existing relevant research most relied on survey data or focused on citizens due to the lack of large-scale, fine-grained tourism datasets. Several scholars have mentioned the notion of trip chains, but few works have been done towards quantitatively identifying the structures of trip chains. In this paper, we propose a model for quantitatively characterizing tourist daily trip chains. After applying this model to tourist mobile phone big data, underlying tourist travel patterns are discovered. Through the framework, we find that: (1) Most "hybrid" (inter-city and intra-city) and "intra-city" (only intra-city) patterns can be captured by only 13 key trip chains relatively; (2) For two continuous days, almost all kinds of original chains have a rather high probability to transfer to either the first two transferred chains, or other infrequent chains in our study areas; (3) The principle of least efforts (PLE) affects tourists' structures of trip chains. We can use average degree and average travel distance to interpret tourist travel behavior (achieving tasks in PLE). This study not only demonstrate the complex daily travel trip chains from tourism big data, but also fill the gap in tourism literature on multi-destination trips by discovering significant and underlying patterns based on mobile datasets.

# 1 Introduction

With the booming development of information and location aware technologies, massive movement datasets can now be obtained to facilitate and improve empirical research on human travel behavior. Many studies have taken such opportunities to quantify human travel patterns, especially from the perspective of trip chains (Csáji et al., 2013; González, Hidalgo, & Barabási, 2008; Jiang, Ferreira, & Gonzalez, 2017; Zhao, Wang, & Deng, 2015). Trip chain is generally regarded as a series of trip segments linked together between two important activity locations (McGuckin & Nakamoto, 2004). Trip chain can reveal how people organize their activities across space and time, as well as the topological relationships among different activity locations. Deeper understanding of trip chains will provide a wealth of information for transportation planners and policy makers, therefore benefit land-use planning and improve urban accessibility, and even forecast and control global spreading of epidemics (Anderson, Anderson, & May, 1992; Hufnagel, Brockmann, & Geisel, 2004; Lloyd & May, 2001; McGuckin & Murakami, 1999; McGuckin, Zmud, & Nakamoto, 2005).

Most prevailing research in trip chaining behavior focuses on activities of city dwellers. It is found that people's travel patterns exhibit a limited number of predictable and characteristic trip chains that follow simple rules (Song, Qu, Blumm, & Barabási, 2010). Namely, several trip chains are sufficient to capture the major characteristics of the populations (Schneider, Belik, Couronne, Smoreda, & Gonzalez, 2013). Besides, citizens' travel decisions are manifested by the principle of least effort --- individuals with different background and socioeconomic characteristics tend to minimize their effort in many aspects of human life (Guy, Curtis, Lin, & Manocha, 2012; Hubbard, 1978; Zipf, 1949). However, few efforts have been devoted to investigating trip chaining behavior of tourists. This is largely due to of the lack of large-scale, fine-grained datasets that capture movement patterns of tourist populations. Note that some models about tourist travel patterns were put forwarded previously (Lau & McKercher,

2006). Despite the fruitful research outcomes, these models are often verified using small-scale datasets such as travel diaries (Mckercher & Lau, 2008). There has been a lack of research on quantifying tourist trip chains using movement datasets with a fine spatiotemporal resolution. Nowadays, analyzing tourists travel patterns becomes significant in helping tourists plan their trips more efficiently. If tourism related business can make some marketing strategies to cater tourists according to the travel patterns, they will develop faster. Therefore, deep understanding of movement patterns regarding tourists from knowledge derived from big data can achieve the ultimate goal --- benefitting both tourists and tourism (Cooper, 1981; Hwang, Gretzel, & Fesenmaier, 2006; Mckercher & Lau, 2008; van der Knaap, 1999), and can be directly applied to the tourism management activities, such as tour product development, attraction planning and accommodation development. From this point of view, this research focuses on mobile phone big data of South Korean tourists and aims to mine tourists' frequent travel patterns.

We are aware that limited quantitative measurements have been put forward when modeling tourists trip chains. Therefore, we focus on addressing following research questions in this study: (1) What are the major topological characteristics of tourist daily trip chains? (2) Do visitors organize their travels in a similar way over different days during their stay in a city? (3) Is the principle of least effort reflected in their travel behavior?

To address these questions, we analyze a large-scale mobile phone dataset collected in two cities in South Korea (Jeonju and Gangneung). The dataset captures the location footprints of international travelers who visited the two cities during a period of one year. The study aims to provide answers to the above questions with a novel model that can capture and reproduce the spatiotemporal structures and regularities in tourist trip chains. In particular, we extract trip chains from raw mobile phone trajectories. After further exploring the most popular ones, typical trip chains of tourists are identified. To be more specific, in order to answer the first research question, we analyze daily trip chains by dividing them into "intra-city" chains and "hybrid" chains. We try to understand how tourists organize their travels during the period that they are within a city, or in the first and last day of their visits to a city. For the second research question, we attempt to find inherent patterns in terms of tourist individual mobility and find out predictability of tourist movement. For the third question, we use two metrics --- average degree and average travel distance of trip chains to understand tourist travel behavior with regard to the principle of least effort.

The rest of the paper is organized as follows. Section 2 reviews the existing literature on mining human mobility patterns. Section 3 presents the study area and dataset used. Section 4 introduce the developed method. Section 5 reports and discuss the results. Finally, Section 6 concludes this study.

## 2 Literature review

A trip involves multiple activities across hierarchical stages of travel experiences (Jeng & Fesenmaier, 2002). That is, a trip is not a simple origin-destination mechanism; rather, it entails multiple destinations (Lue, Crompton, & Fesenmaier, 1993). To explicate patterns associated with these multi-trips, tourism scholars have proposed various approaches, from spatial configuration (Lue et al., 1993; Lew & McKercher, 2006) to activity-based perspectives (Woodside & Dubelaar, 2002; Woodside & King, 2001). These relevant studies argue that it is fallacious to assume travelers go to a single place (or destination) after leaving home. Individuals show multi-destination patterns in which they make numerous activity decisions that influence behavior in an interactive fashion. The pattern of multi-destination trips is associated with the notion of a rational behavior wherein individuals are likely to minimize time and cost associated with travel, an effect that potentially increases accrual of benefits and fulfills the desire for variety in destinations.

In this sense, the current study suggests the concept of trip chain to elucidate multi-destination trip patterns, as have been widely discussed in the literature on human mobility and transportation (see Primerano, Taylor, Pitaksringkarn, & Tisato, 2008). The notion of trip chain varies depending on the various contexts in which it has been applied (Goulias & Kitamura, 1991; Kumar & Levinson, 1995; McGuckin & Murakami, 1999). From the viewpoint of transportation, Primerano et al., (2008) summarized two commonly used definitions: (1) "A sequence of trip segments begins at the home activity and ends when the individual returns home"; (2) "A sequence of trip segments between a pair of activities: home, work, or school" (page 58). The trip chain approach enables researchers to identify similarity/regularity of public transit patterns (Ma et al., 2013), design activity schedule linking primary (e.g., home) and secondary (e.g., work) activities (Primerano et al., 2008), and compare different usages of transportation facilities according to socio-demographic features (McGuckin & Murakami, 1999; Zhao, Wang, & Deng, 2015). According to human mobility, trip chain can be regarded as a mobility network in essence whereby the activity (or a place visited) can be denoted as a node and the spatial pattern/links (or flow) can be indicated as edges. In this sense, another similar concept mobility motif, which refers to highly-repeated multidimensional subsequences in a complex mobility network structure, is also proposed to clarify the mobility patterns (Thomas F Golob, 2000; Holguín-Veras & Patil, 2005; Maruyama & Harata, 2006; Murray-Tuite & Mahmassani, 2003; Zhang, Baohua, Mingjun, Jinchuan, & Jifu, 2007). For instance, Schneider et al. (2013) described a kind of individual daily movement network as a mobility motif if it occurred more than 0.5% in the datasets. Indeed, while different names were used (e.g., "trip chain" and "mobility motif"), these consistently represent methods to interpret individuals' travel patterns by distinguishing between locations. Principally, "trip chain" refers to complex relationships between a set of activities and the interdependence of temporal (e.g., timing, duration, length, and sequence of trips) and spatial (e.g., location) characteristics associated with human mobility. Thus, the trip chain model in tourism can be defined as a sequential pattern of trip activities (or places visited for travel activities) made by travelers on a day-to-day basis (Golob & Hensher, 2007; McGuckin, Zmud, & Nakamoto, 2005).

Tourism scholars have applied the notion of trip chain as a means of comprehending travel movement behaviors. Lue et al. (1993), in a pioneering study, conceptually and primarily suggested a trip-chaining pattern as a type of spatial model of pleasure vacation trips, illustrating visitations involving numerous focal activities. Along with their own work, several tourism scholars (e.g., Stewart & Vogt, 1996) tried to demonstrate the trip chain patterns in their own terms but on consistent ideas. For example, Lau and McKercher (2006) presented a chaining loop as part of tourist movement patterns describing a certain pattern of visiting multiple destinations. Likewise, Lew and McKercher (2006) proposed conceptual linear path models of tourist behavior in intra-destinations consisting of point-to-point patterns (Type I), circular patterns (Type II), and complex patterns (Type III). The previous tourism literature, in theory, has discussed the notion of trip chains and its application to deconstruct travel movement patterns.

Importantly, however, most studies contain the limitation with lack of quantitative verifications of the models. This may be attributable to challenges faced in accessing tourism big data that provides comprehensive insights into the phenomenon and can be used to calibrate the models. The tourism studies typically have collected data on travel behaviors using surveys. Such approaches typically require a substantial financial outlay and expenditure of effort. They also contain the potential for response errors such as cognitive bias of respondents (Shoval & Ahas, 2016). With the evolution of information technology, a number of tourism researchers have adopted social media data such as Flicker, Twitter, and Weibo (Arase, Xie, Hara, & Nishio, 2010; Paldino, Bojic, Sobolevsky, Ratti, & González, 2015; Yang, Wu, Liu, & Kang, 2017; Zeng, Zhang, Liu, Guo, & Sun, 2012). Nevertheless, analysis of social media content suffers from sparseness of data, making it difficult to discern the comprehensive travel patterns. Thus, our study applies a trip-chaining method to tourist mobile phone data, enabling tourism researchers to overwhelm the shortcomings from traditional data and to uncover hidden patterns, which ultimately discover underlying spatial behaviors of tourists.

Travel mobility is closely related to travel distance that reflects individual efforts being consumed to reach his/her goals (e.g., arrival to a place; Gonzalez, Hidalgo, & Barabasi, 2008). A broad theory of the principle of least effort (PLE) proposed by George Kingsley Zipf (1949) supports this argument. The model claims that people tend to choose the method requiring the least effort to finish tasks. As an example, Zipf discovered a certain speech pattern in which people tend to use short words for their daily communication. That is, the distribution between word frequency used by speakers and hearers and word rank is largely skewed and demonstrated by a mathematical formula, now called Zipf's law (Manning & Schutze, 1999). Applying PLE to travel mobility, the task encompasses movement; travelers would likely to seek out the "optimal way" to minimize the total movement required. It can be thus argued that travel distance is related to different types of trip chain models. Other than understanding of human mobility, the PLE has been widely applied to explain a variety of human behaviors including information-seeking behavior (Baruchson-Arbib & Bronstein, 2007; Chang, 2016), human mobility (Cao, Li, Tu, & Wang, 2019), pedestrian mobility (Guy et al., 2010), and street networks (Masucci, Smith, Crooks, Batty, 2009).

## 3 Study area and dataset

South Korea is a country with a well-developed travel & tourism industry. Jeonju and Gangneung, which are two popular cities to international travelers in South Korea, are selected as areas of study. Jeonju, the capital city of Jeollabuk-do Province, is an important tourist center famous for traditional Korean food, historic buildings, sports activities, and festivals. It has an area of 206 km$^2$ and a population of 0.65 million (as of 2017). Gangneung sits on the east coast of South Korea. The city has many tourist attractions, such as Jeongdongjin, a very popular area for watching the sun rise, and Gyeongpo Beach. It's also the city that hosted all the ice events for the 2018 Winter

Olympics. As a city in Gangwon-do Province, it has an area of 1040 km$^2$ and a population of 0.21 million (as of 2019).

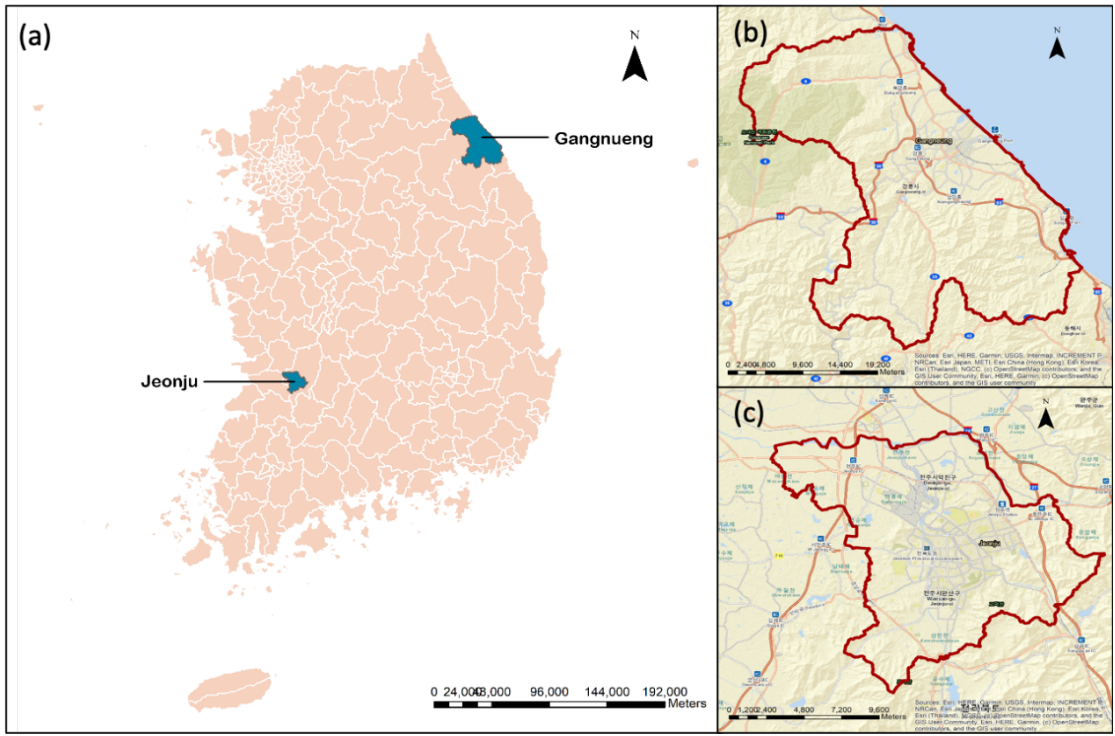

Figure 1. Study areas: (a) the whole South Korean, (b) Gangnueng, and (c) Jeonju.

This research uses a mobile roaming dataset collected by a major telecom company in South Korea. This anonymized dataset tracks the location footprints of 18,625 and 33,219 tourists who visited Jeonju and Gangneung respectively between August 1$^{th}$, 2017 to July 31$^{st}$, 2018. Since the timespan of dataset covers the 2018 Winter Olympics and mobility patterns of tourists could be different during this special event, we filter out this part of the data (from Jan 20$^{th}$, 2018 to Feb 26$^{th}$, 2018). Thus, the number of tourists in Gangneung changes to 15,095. Note that in this dataset, as long as an individual has visited Jeonju or Gangneung, the sequence of locations that he or she stayed when travelling in any other city of South Korea were also documented. This reveals additional information on when an individual entered/left Jeonju or Gangneung, which enables precise quantification of his/her trip chain on the first or last day of visit to a city. Table 1 shows an example of an individual's phone records. Each row in the table represents one stay activity and the time periods in between indicate trips among locations. For example, the first two rows in Table 1 indicate that the user stayed at two different locations during [07:16:00 - 12:33:00] and [12:46:00 – 12:52:00] respectively, and a trip was possibly conducted by the user in between (i.e., [12:33:00 - 12:46:00]).

Table 1. Example of an individual's mobile phone records

| User ID | Date | Starting Time | Ending Time | Longitude | Latitude |
|---|---|---|---|---|---|
| 28*** | 2017-08-25 | 07:16:00 | 12:33:00 | 127.*** | 35.*** |
| 28*** | 2017-08-25 | 12:46:00 | 12:52:00 | 127.*** | 36.*** |
| 28*** | 2017-08-25 | 13:08:00 | 13:24:00 | 127.*** | 36.*** |
| … | … | … | … | … | … |
| 28*** | 2017-09-06 | 15:01:00 | 15:14:00 | 126.*** | 35.*** |
| 28*** | 2017-09-06 | 15:43:00 | 16:07:00 | 126.*** | 35.*** |
| 28*** | 2017-09-06 | 16:41:00 | 17:00:00 | 126.*** | 35.*** |

The locations of users were positioned at the level of cellphone tower and their densities in space define the spatial resolution of the dataset. The numbers of cellphone towers in Jeonju and Gangneung are 782 and 704, respectively. To better understand their spatial arrangement in the two cities, we calculate the statistics of the distance from each cellphone tower to its nearest peer. The average distance is 250 meters in Jeonju and 420 meters in Gangneung. Overall, the dataset provides a fine-grained view of tourist mobility in time and space, which allows for reliable extraction of tourist trip chains.

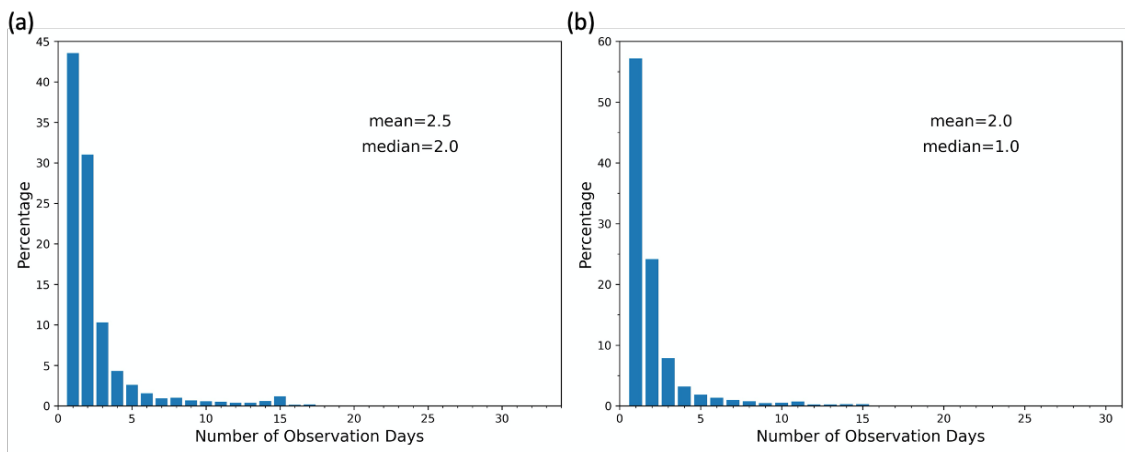

Figure 2. The number of observation days of users in two cities.

As shown in Figure 2, the number of observation days of users (i.e., the number of days with records for a user) show similar distribution patterns in two cities. Most of the users stayed for only a few days. The median values of observation days are 2.0 for Jeonju and 1.0 for Gangneung, and the mean values are 2.5 for Jeonju and 2.0 for Gangneung. The maximum values of observation days are 34 days for Jeonju and 30 days for Gangneung, respectively. Note that for some travellers, they would stay in a particular city for a few days and left, and then came back for another visit. In this study, we don't perform trip chain analysis for individuals with gap days. Such individuals account for 7.22% and 12.66% of visitors in Jeonju and Gangneung, respectively.

# 4 Methods

To understand and quantitatively model tourists' mobility patterns, we design a pipeline to analyze the trip chains of tourists. It contains four steps: (1) extract meaningful activity locations ("anchor points") from tourists' cellphone trajectories; (2) construct daily trip chains of tourists and examine their key characteristics; (3) quantify the day-to-day transitions of individual trip chains; (4) explore decisive factors that shape tourists' trip chains.

## 4.1 Derive individual activity anchor points from cellphone trajectories

Identifying meaningful locations of travellers is an essential step for trip chain analysis. However, the cellphone tower locations documented in the dataset do not always reflect the actual locations of users for several main reasons: (1) the signal of mobile phones could switch between adjacent cellphone towers, producing the so-called "ping-pong" effect (Inzerilli, Vegni, Neri, & Cusani, 2008; Rasmussen & Oppermann, 2003). (2) the signal transmitted and received by cellphone towers will be compromised during propagation, leading to the inaccuracy of cellphone tower positioning (Isaacman et al., 2012). Therefore, we argue that the combination of neighboring cellphone towers could better represent a location or place that is meaningful to a traveler.

Activity anchor points have been used in previous studies to describe a person's major activity locations (Ahas, Silm, Järv, Saluveer, & Tiru, 2010; Dijst, 1999; Schönfelder & Axhausen, 2003). In this study, we define an *anchor point (AP)* as a set of cellphone towers that are close to each other and where an individual has stayed over a certain period of time.

Given a cellphone trajectory $T = \{(v_1, t_1^s, t_1^e), (v_2, t_2^s, t_2^e), \ldots, (v_n, t_n^s, t_n^e)\}$ that documents a user's location footprints. Here $v_i = (lng_i, lat_i)$ symbolizes the cellphone tower location of the $i^{th}$ record; $t_i^s$ and $t_i^e$ represent the starting and ending time of $i^{th}$ record. The anchor point extraction works as follows. First, we calculate the total amount of time the individual stayed at each cellphone tower, and sort all the cellphone tower in descending order based on total stay duration. We start from the cellphone tower with the largest stay duration and group other cellphone towers within a roaming distance ($\Delta d$) of the selected cellphone tower into a cluster. Then from left cellphone towers that have not been assigned to any cluster, we conduct the same process on the one with longest stay duration. Iterating above steps until all the cellphone towers in trace $T$ are tackled with, the trajectory is processed into sequence at anchor point level. Since the average nearest distance between cellphone towers in both Jeonju and Gangneung are below 500m, we set $\Delta d$ as 500m for both cities when performing the anchor point extraction.

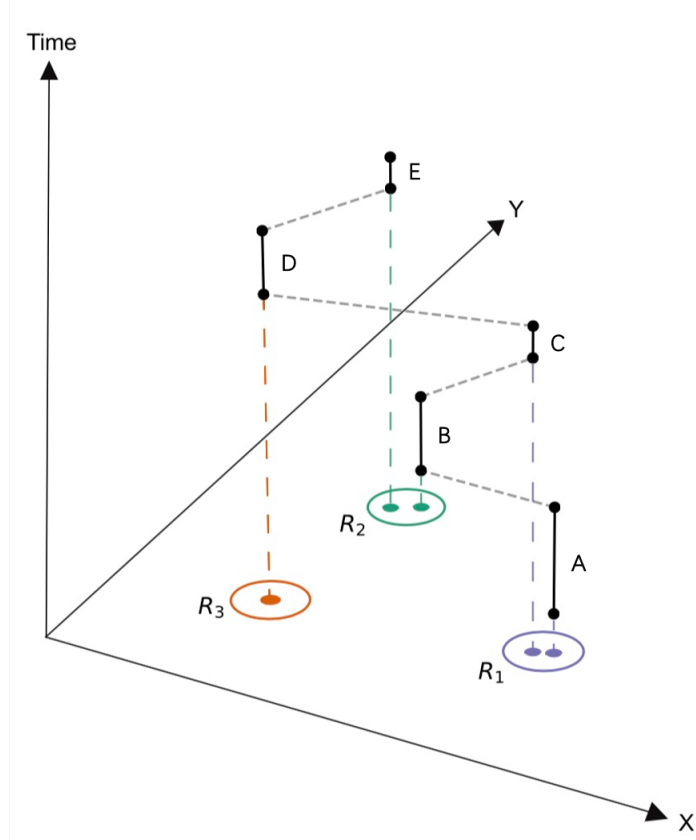

Figure 3. An example of extracting anchor points from a trajectory. The five vertical line segments mean five rows of records. $A$ to $E$ represent five cellphone towers in the trajectory $T$. $R_1$ to $R_3$ denote the extracted activity anchor points.

In this way, an individual's trajectory can be represented as a sequence of APs. Figure 3 displays an example of how APs are extracted from the cellphone trajectory. Given the individual's cellphone trace $T = \{(A, t_A^s, t_A^e), (B, t_B^s, t_B^e), \ldots, (E, t_E^s, t_E^e)\}$, the travel patterns can be denoted as a sequence of cellphone tower locations traversed: $A \rightarrow B \rightarrow C \rightarrow D \rightarrow E$. In order to implement the above workflow, the cellphone tower with the longest total duration ($A$) is selected to create the buffer. The distance between cellphone tower $C$ and $A$ is less than 500m, so they are grouped as AP $R_1$. Then, the next cellphone tower with the highest amount of time ($B$) is selected and grouped with $E$ to form $R_2$. The left cellphone tower $D$ form AP $R_3$ itself. At this point, the individual's travel patterns can be redefined as a sequence of APs: $R_1 \rightarrow R_2 \rightarrow R_1 \rightarrow R_3 \rightarrow R_2$. And the trace is processed to $T' = \{(R_1, t_{R_1}^s, t_{R_1}^e), (R_2, t_{R_2}^s, t_{R_2}^e), \ldots, (R_2, t_{R_2}^s, t_{R_2}^e)\}$, where $R_j = (lng_j, lat_j)$ denotes the location and ID of the $j^{th}$ AP. So far, we have converted original observations at cellphone tower level to sequence at AP level.

## 4.2 Construct tourist daily trip chains

To further picture tourist daily travel patterns, we construct trip chains with different topological structures, which reveal how tourists organize their daily travels. Different from previous human mobility research which focuses on residents (Alessandretti, Sapiezynski, Sekara, Lehmann, & Baronchelli, 2018; Csáji et al., 2013; Jiang et al., 2017; Schneider et al., 2013), our research objects are tourists in this study. Residents' movements are usually within a given city; while for those tourists who visited several cities in one travel, they have records out of the given city. Thus, when we construct daily trip chains for a cross-city traveler, we have two kinds of trip chains: (1) "hybrid" trip chains, which are composed of all records of an individual including within-city and out-of-city observations, and (2) "intra-city" trip chains, which refer to trip chains constructed with only intra-city records. For the "hybrid" trip chains, since the purpose of keeping the out of city records is to explore the travel pattern features of the day when cross-city travelers come or leave a given city, we only need to count the relevant records before arriving or after leaving the city as one aggregated AP respectively.

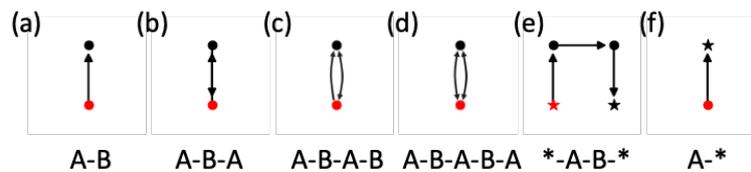

Figure 4. Examples of daily trip chain networks. (a) denotes an individual with daily trip chain from anchor A to anchor B (A-B). (b) denotes an individual with daily trip chain from anchor A to anchor B then to anchor A (A-B-A). (c) denotes an individual with daily trip chain A-B-A-B. (d) denotes an individual with daily trip chain A-B-A-B-A. (e) denotes an individual with daily trip chain *-A-B-*, he or she is out of city at the beginning and end, and visited two "within city" anchors in the process. e) denotes an individual with daily trip chain A-*, he or she stays in "within city" anchor A at the beginning, then left the city to an "out of city" anchor.

As shown in Figure 4, when we plot the structure of trip chains, we use an edge with a one-way arrow to represent an individual from an anchor to another. We use an edge with a two-way arrow to represent a round trip between two APs. Red symbol represents the start anchor. Point in round shape denotes anchor points within city. Point in star shape denotes anchor points out of city. We use "*" to symbolize "out of city" anchors, and capital letter to symbolize "within city" anchors. Figure 4 (a) - (d) show typical "intra-city" trip chains, (e) and (f) show typical "hybrid" trip chains.

## 4.3 Day-to-day transition of tourist trip chains

To understand how an individual changes his or her travel patterns on the base of consecutive days, we need to analyze transition between continuous daily trip chains. As we have mentioned in Figure 2, the median values of observation days are 2.0 for both cities. Thus, it's meaningful to explore the transition patterns of tourists' daily mobility between two consecutive days. In this part, we focus specifically on whether individual will change his or her type of travel chain on the base of consecutive sequence. By doing so, inherent pattern in tourist individual mobility will be explored, which may provide new insights into tourist mobility.

Note that days coming or leaving the given city are not comparable to days staying in the city, we only focus on "intra-city" trip chains in this section. In this part, only individuals with at least two consecutive days of "intra-city" trips are analyzed. Suppose a tourist has records for $Q$ days of "intra-city" trips, we can then extract $Q-1$ pairs of consecutive days. For each pair of consecutive days, we define the trip chain of the former day as original chain, and the trip chain of the latter day as transferred chain. A transition matrix will then be constructed to count the frequency of different combinations of trip chain transition. Row indexes of the transition matrix are different types of original chains, and columns indexes are different types of transferred chains. We first investigate the frequency value of each element in the transition matrix, then we turn frequency into probability by summing each row. By doing so, each element value in the matrix presents the likelihood that people convert one type of original chain to different types of transferred chains. For instance, suppose a person has records in a given city for $W$ continuous days ($W$ is not less than 2). He or she stayed in one place

(chain type: A) for whole day the first day, and conducted a round trip with two nodes A,B (chain type: A-B-A) the second day, then the frequency of the matrix element corresponding to original chain A to transferred chain A-B-A should add one. Then we count the frequency for transition of other $W-2$ pairs of consecutive days of this individual. After all tourists in the given city are performed above procedures, the probability value of one original chain type changing to others transferred chain types can be calculated.

## 4.4 The principle of least effort in trip chaining behavior

We argue that the principle of least effort also acts as driving force in trip chaining behavior. In order to demonstrate it is a decisive factor in tourist travel behavior, we need to adopt some statistical metrics. For example, distance is a key metric in human mobility (Olsson, 1965), and average degree is an important metric in network-like topological structure. Thus, in this paper, we use two indicators to represent two aspects of trip chaining behavior: (1) average degree $K$, and (2) average distance $D$. These two indicators are defined and calculated as follows:

$$K = \frac{E}{N} \quad (1)$$

$$D = \frac{\sum_{i=1}^{N-1} d_{i,i+1}}{E} \quad (2)$$

Where $E$ is the edge number of each trip chain, and $N$ represents the AP number (we denote AP number as $N$ afterwards). $i$ and $i+1$ are a pair of consecutive APs in a chain, and $d_{i,i+1}$ represents Euclidian distance between AP $i$ and AP $i+1$.

The propose of investigating average degree $K$ is to explore whether there is some connection between travel efficiency and people's preference of choosing daily trip chains. $K$ can be regarded as a proxy of travel efficiency. To be more specific, through the network structure of trip chains, we can clearly know whether people prefer to visit different locations in a single round tour before returning to the starting location, or if they prefer to return to their starting location before visiting another location. Obviously, the most effective way to conduct an itinerary with $M$ APs is a round trip with $M$ segments, in which case, $K$ is equal to 1. If a person moves multiple times between APs, then $K$ is larger than 1. The higher the value of $K$, the less efficiency the individual travel is.

As for average distance $D$, it makes sense that people usually think ahead about the next day's itinerary, including how many places they want to go and distances of these places in space. We calculate average distance for each sample, to combine the information of AP number $N$ and Euclidian distance between two consecutive APs in the trip chain. To some extent, the average travel distance of a trip chain can be considered as an integration of multiple factors related to the principle of least effort when people choose trip chains, such as travel cost, spatial range and mobility regularity. Average distance can reflect cost people pay for travel in terms of time and space. From this point of view, the travel distance metric can offer another standpoint towards understanding tourist travel behaviors.

# 5  Results and discussions

## 5.1 Distribution patterns of anchor points

In this section, we report the distribution of the number of activity anchor points extracted from tourists' cellphone trajectories in Jeonju and Gangneung. We then explore the spatial patterns of these anchor points to gain insights into the tourists' spatial preferences in the two cities.

**Numerical distribution of anchor points**

We first investigate the number of daily visited AP on individual basis. Through the numerical distribution of AP, we can have an insight into one aspect of tourisms' daily travel preference.

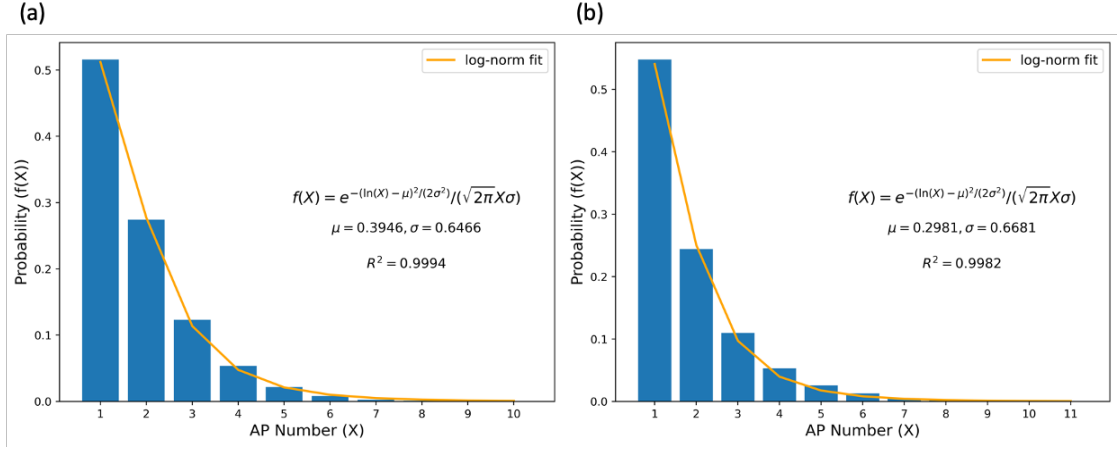

Figure 5. Distributions of AP number for (a) Jeonju dataset, (b) Gangneung dataset. Tourists in Jeonju and Gangneung visited up to 10 and 11 APs in a daily basis respectively. The values of blue bars are the observed probabilities of different number of nodes. The yellow lines denote the best-fitted distributions --- log-normal. The fitting was conducted via the least squares fit. The corresponding function and parameters are shown in each figure.

The distribution of AP as well as the fitting of the probability of travelers who visited a given AP number is presented in Figure 5. The probabilities $P$ of different AP number are: $P_{Jeonju}(X=1) = 0.52$, $P_{Jeonju}(X=2) = 0.27$, $P_{Jeonju}(X=3) = 0.12$; $P_{Gangneung}(X=1) = 0.55$, $P_{Gangneung}(X=2) = 0.24$, $P_{Gangneung}(X=3) = 0.11$. In addition, it's seems that daily human mobility patterns follow a universal law. The number of daily visited APs can be approximated with a log-normal fit:

$$f(X) = \frac{e^{\frac{-(\ln X - \mu)^2}{2\sigma^2}}}{\sqrt{2}X\sigma} \tag{3}$$

with the parameters $\mu_{Jeonju} = 0.3946$, $\sigma_{Jeonju} = 0.6466$, $\mu_{Gangneung} = 0.2981$, $\sigma_{Gangneung} = 0.6681$.

The parameters are calculated with 95% confidence bounds, and R-squares of fittings are both above 0.99. The two datasets demonstrate great internal heterogeneity through log-normal fitting, which indicates travelers are diverse in terms of the number of places visited in a day. However, the similar values of $\mu$ and $\sigma$ in two datasets reveal that distributions extracted from Jeonju and Gangneung dataset show similar patterns. In other words, the variance in travelers' spatial behavior are comparable between the two cities. The average AP number $\hat{X} \approx 2$ (1.83 for Jeonju and 1.84 for Gangneung) is small; hence, most people visit only a few places. In fact, 99.60 / 99.24 per cent of the population visit less than seven APs on a daily basis in Jeonju / Gangneung.

The tail of the distributions shows that although most people visit less than four APs, a small fraction of tourists visit quite a lot APs within a day in some cities.

**Spatial patterns of tourist activities**

In order to explore the hot spots of our study areas through meaningful activity anchor points, we use kernel density method to plot heatmap of anchors. A smaller radius of kernel density can make the local patterns obvious, while a larger radius can generate a smoother global surface. Note that the radiuses for kernel density need to be larger than the average distance of cellphone towers (250m in Jeonju and 420m in Gangneung). We also need to take the area of cities into account (206.22 km² for Jeonju and 1040 km² for Gangneung). Since our goal is to identify hotspots by kernel density, the search radiuses for Jeonju and Gangneung are set as 1000m and 3000m respectively.

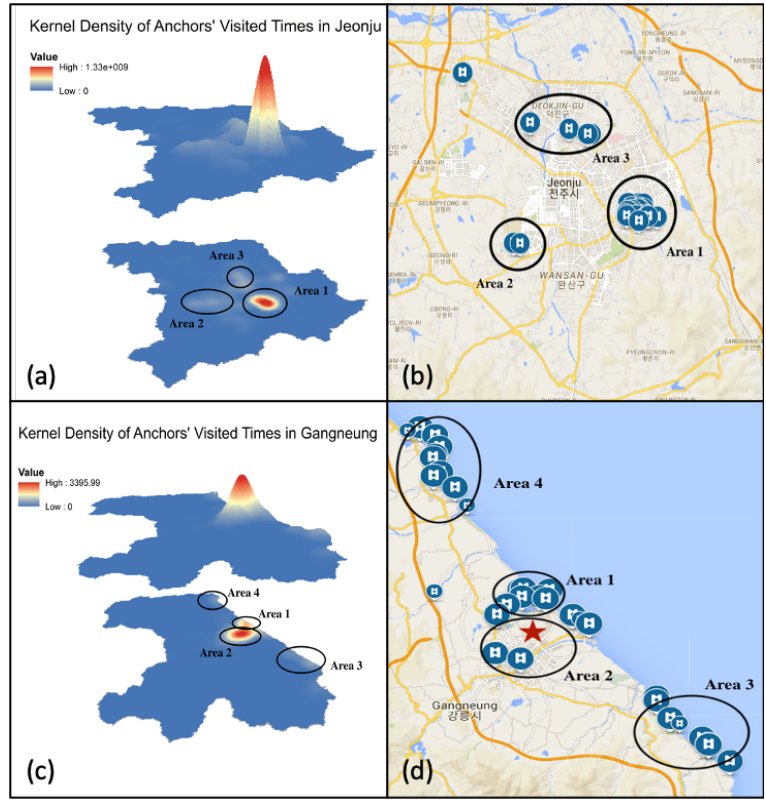

Figure 6. Popular areas derived from datasets (a) Jeonju (c) Gangneung, and hot spots marked by TripAdvisor (b) spots in Jeonju (d) Gangneung. In (a) (c), popular areas are derived through visited times of AP. The deeper the blue, the lower the value of kernel density; the deeper the red, the higher the value of kernel density. In (b) (d), the blue marks correspond to locations of spots in Jeonju and Gangneung respectively. The red star area in (d) represents Winter Olympic venues. Area 1, 2, 3 in (a) has the same locations with Area 1, 2, 3 in (b); Area 1, 2, 3, 4 in (c) has the same locations with Area 1, 2, 3, 4 in (d).

As shown in Figure 6, even though some areas of spots are not detected, the distributions of popular areas of two cities derived by kernel density generally match with with the ground truth provided by TripAdvisor. Note that the kernel density maps are based on one year's mobile phone data, while maps of TripAdvisor are based on Points of Interests. Thus, this section can be seen as a validation of our deriving activity anchor point method. Moreover, results of kernel density also imply which areas are popular in 2017-2018.

## 5.2 Significant trip chain types

In order to discover the most popular ways for tourists in South Korea to organize their daily travels, we construct significant trip chains and visualize them. To keep it consistent, for "hybrid" trip chains and "intra-city" trip chains, we regard the top 13 trip chains among each of them as significant chain types in this research, since they all account for more than 1% of the total trip chains. in the respective analysis.

**Significant "hybrid" trip chains**

We argue that if we extend trip chains to "hybrid" perspective, more patterns about days leaving or coming to the given city can be observed.

Figure 7 shows the top 13 "hybrid" chains, which include both individual trip chains which are all in the city and individual trip chains which are sometimes out of city. This would reveal the behavioral diversity of travelers on an average day. To summarize, up to 76.41% and 75.99% of the measured "hybrid" trip chain types can be described with only 13 different daily trip chains in Jeonju and Gangneung respectively. In general, the trip chains can be grouped into four main categories based on the start nodes and end nodes:

**(C1) staying in the city:** the start point and end point of the trip chain are both in round shape, which means the user were in the city at the beginning and the end. Or the motif just has one single point, which means the user spent all day in one anchor point;

**(C2) passing by the city:** the start point and end point of the trip chain are both in star shape, which means the user were out of city at the beginning and the end;

**(C3) coming to the city:** the start point of the trip chain is in star shape while the end point of the trip chain is in round shape, which means the user were in the city at the beginning and out of city at the end;

**(C4) leaving the city:** the start point of the trip chain is in round shape while the end point of the trip chain is in star shape, which means the user were out of the city at the beginning and in the city at the end.

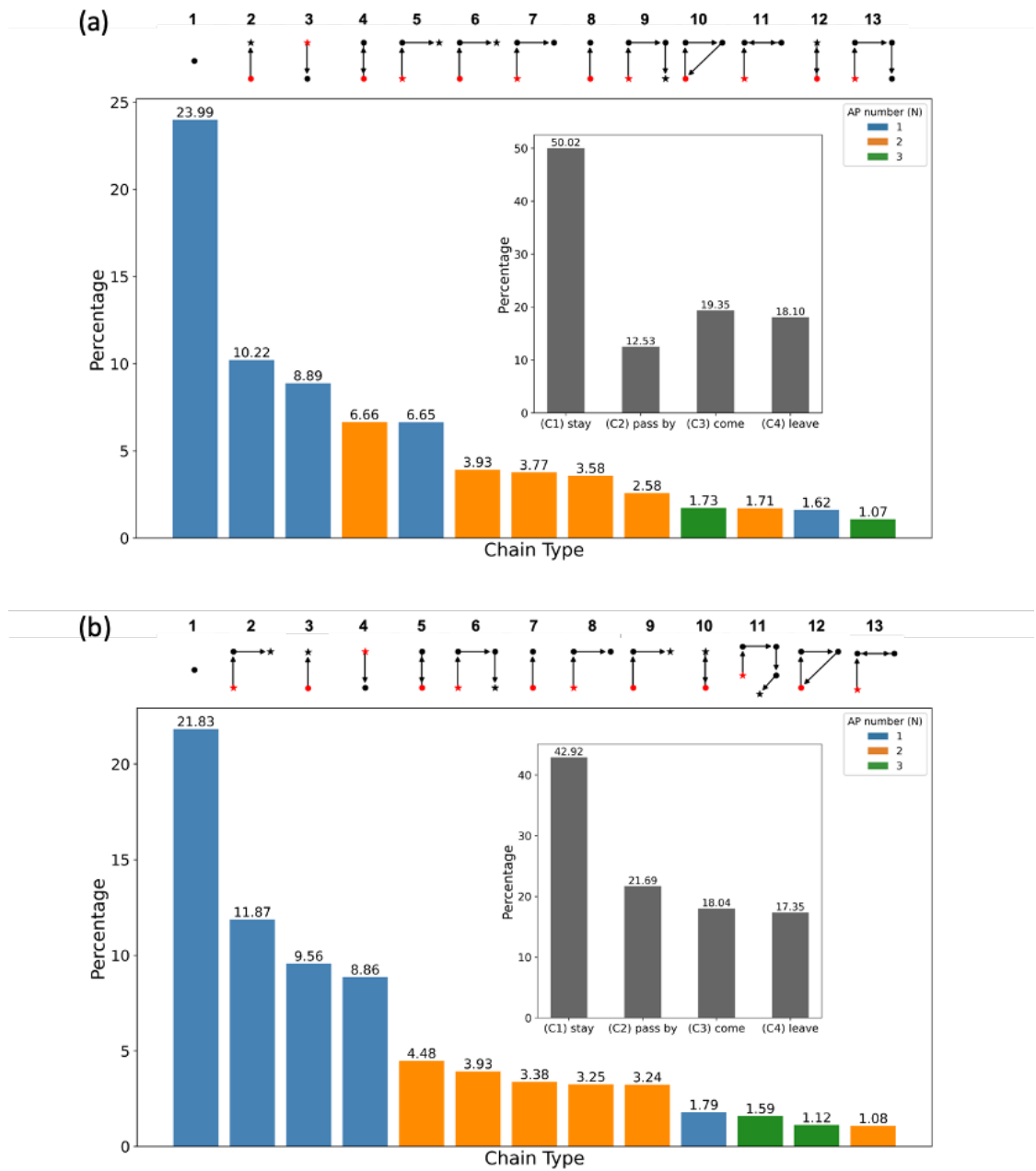

Figure 7. Significant "hybrid" trip chain types in (a) Jeonju, (b) Gangneung. There are 1866 different chain types in Jeonju and 1655 in Gangneung. The different colors of bar indicate the number of "within city" AP in a chain. The topological structures and their ID (1-13) are shown at the top. In addition, we divide tourists into four categories: (C1) staying in the city, (C2) passing by the city, (C3) coming to the city, (C4) leaving the city. The small bar graphs with grey bars show the proportion of them.

We can see from the Figure 7 that, about half of individual daily trip chains belong to the category of staying in two cities; the amount of individual daily trip chains of coming to the city and leaving the city is almost same in these two study areas; the amount of individual daily trip chains of passing by Gangneung is larger than that of Jeonju, accounting for 21.69% and 12.53% respectively. Chain type 3 in Jeonju and chain type 4 in Gangneung indicate for most tourists who come to the given cities, they prefer to go directly to a place (e.g., hotel) and stayed thereafter the first day. Chain type 5 in Jeonju and chain type 2 in Gangneung indicate most tourists who pass by the given cities only visit one spot.

**Significant "intra-city" trip chains**

Figure 7 suggests that the category *(C1): staying in the city* accounts for about half of samples. In this section, we focus on this category and analyze significant "intra-city" trip chains, to discover the most popular patterns for tourists at the intra-city level.

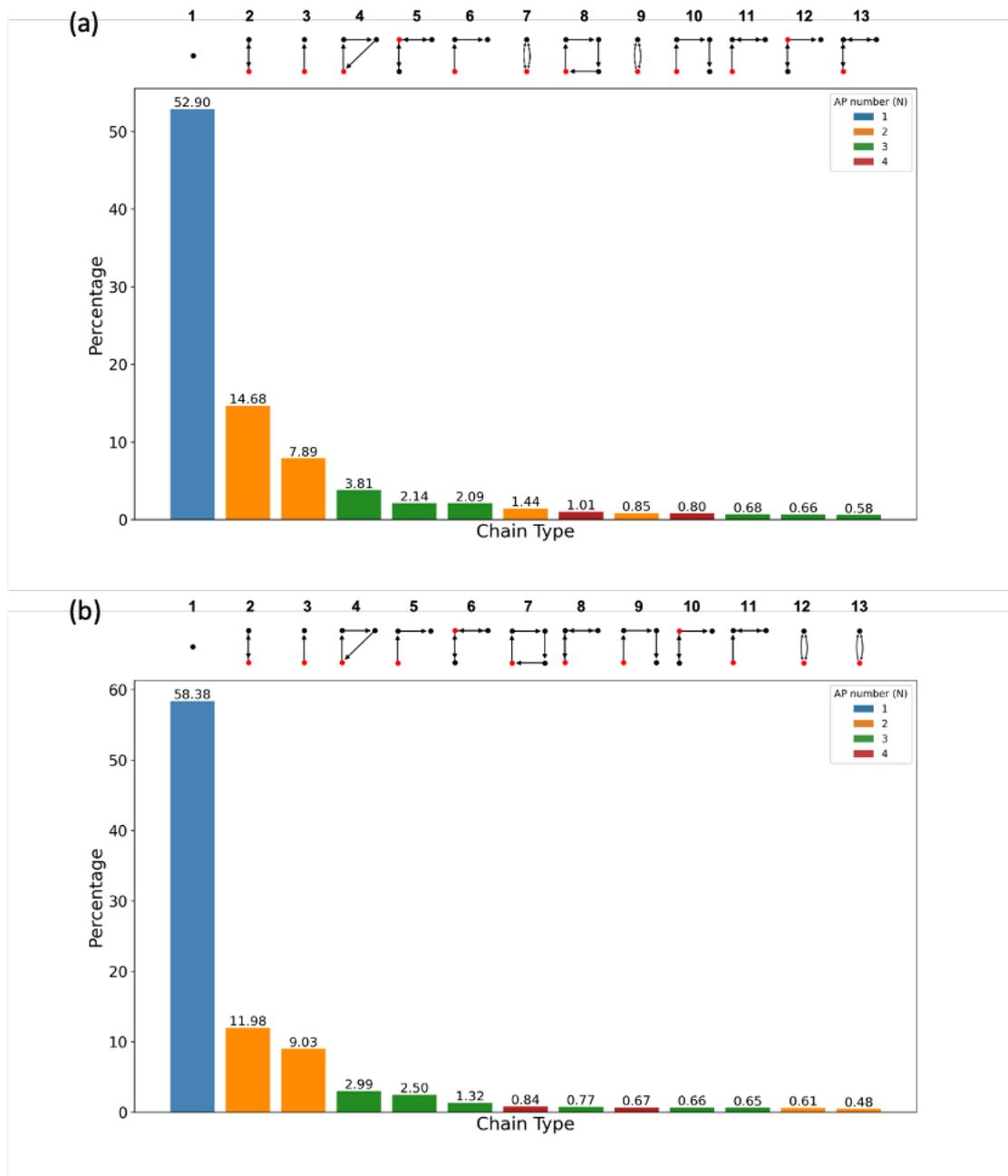

Figure 8. Significant "intra-city" trip chain types in (a) Jeonju, (b) Gangneung. There are 743 different chain types in Jeonju and 467 in Gangneung. The different colors of bar indicate the number of "within city" AP in a chain. The topological structures and their ID (1-13) are shown at the top.

The distribution of intra-city trip chain type samples is displayed in Figure 8. 89.55% and 90.89% of the measured intra-city chain types can be identified with 13 different daily networks in Jeonju and Gangneung respectively. And the first three kind of chain types (ID 1,2,3) account for a quite large part (75.47% for Jeonju and 68.13% for Gangneung) of all chain types. Due to the phone positioning mechanism of roaming dataset, if a person keeps moving in a period, then no record will be documented during this period. Thus, the first three kind of chain types correspond to several travel behaviors such as staying at hotel all day, hanging out near the hotel, and visiting very limited places within a day. It implies tourists tend to conduct quite simple daily tours in intra-city tourism. In addition, the similarity of results in two cities also demonstrates the intrinsic properties of tourist mobility. These findings may provide some insights for tourism planning. For example, most tourists in these two cities may prefer integrated attractions rather than decentralized and monotonous ones.

## 5.3 Day-to-day transition of tourist trip chains

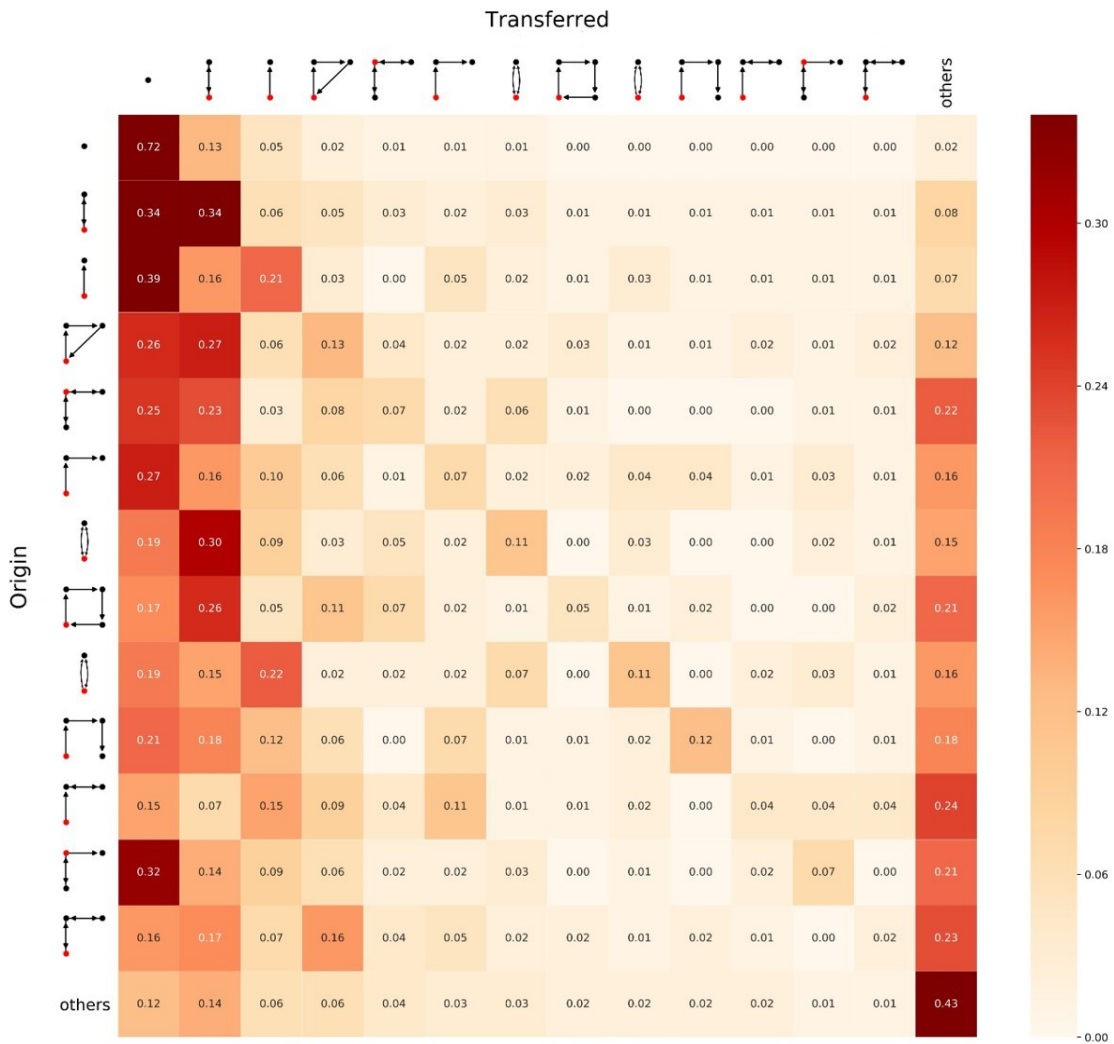

(a) Jeonju

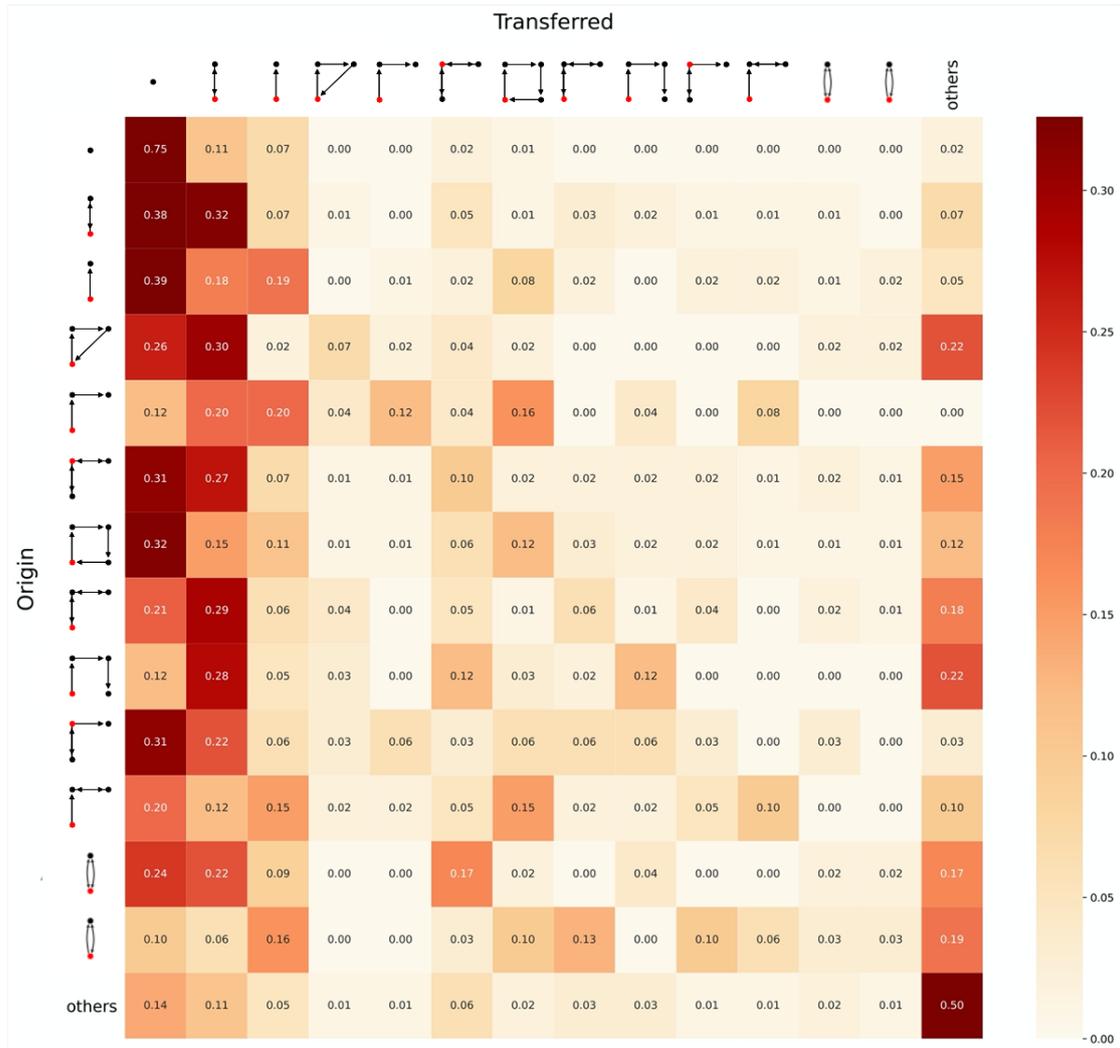

(b) Gangneung

Figure 9. Transition matrix for (a) Jeonju dataset, (b) Gangneung dataset. Every row represents a kind of trip chain that individuals conduct in the first day, every column represents a kind of trip chain that individuals conduct in the second day. For trip chains that are not significant, we conclude them into "others".

Figure 9 shows the transition matrix for two cities. We can see that the transition daily mobility patterns of consecutive sequence-based chains in two cities are quite similar. The values, which equal to the probabilities of corresponding two kind of trip chains in two continuous days, show how more or less likely an original travel chain transfers to another under the condition that the individual has mobility on two continuous days. The darker the colors, the higher the values. By locating grids with dark color, it can be found that almost all kinds of original chains have a rather high probability to transfer to either the first two transferred chains, or "others", in both Jeonju and Gangneung.

The emerging patterns of transitions could be interpreted in two aspects. First, Jeonju and Gangneung are cities with not many tourist attractions since the median values of observation days of tourists are 2 for both cities. For some tourists who just pass by these cities, they prefer to stay in hotel to get fully rest, or go to one spot for short sightseeing or tasting local food. For other tourists who want to explore these cities, they may also like to take one day to get fully rest in next day after they have a whole-day sightseeing. Second, the high probability of transferring to "other" maybe related to long tail principles. The type "others" consists of 730 chain types in Jeonju and 454 chain types in Gangneung, when all the situations are accumulated, the probability of significant original chain types transferring to "others" will become relatively high.

Table 2. The probability of transition for chains with different AP number in Jeonju

| Transferred / Origin | N = 1 | N = 2 | N = 3 | N = 4 |
|---|---|---|---|---|
| N = 1 | **0.72** | 0.19 | 0.06 | 0.01 |
| N = 2 | 0.33 | **0.44** | 0.12 | 0.02 |
| N = 3 | 0.25 | **0.32** | 0.22 | 0.04 |
| N = 4 | 0.19 | **0.33** | 0.20 | 0.10 |

Table 3. The probability of transition for chains with different AP number in Gangneung

| Transferred / Origin | N = 1 | N = 2 | N = 3 | N = 4 |
|---|---|---|---|---|
| N = 1 | **0.75** | 0.18 | 0.04 | 0.01 |
| N = 2 | 0.37 | **0.41** | 0.14 | 0.02 |
| N = 3 | 0.27 | **0.34** | 0.23 | 0.02 |
| N = 4 | 0.18 | 0.27 | **0.31** | 0.05 |

When we aggregate the chains by AP number $N$, we can clearly see from Table 2 and Table 3 that for top 13 significant chains in each city, there is a high probability for them transferring to chains with one or two AP. The results further demonstrate our above conclusions.

## 5.4 The principle of least effort in trip chaining behavior

As mentioned in section 4.4, we hypothesize that travel efficiency is the substantial factor in tourist travel behavior. We checked the proxy of travel efficiency, chain degree $K$, in order to validate our assumption.

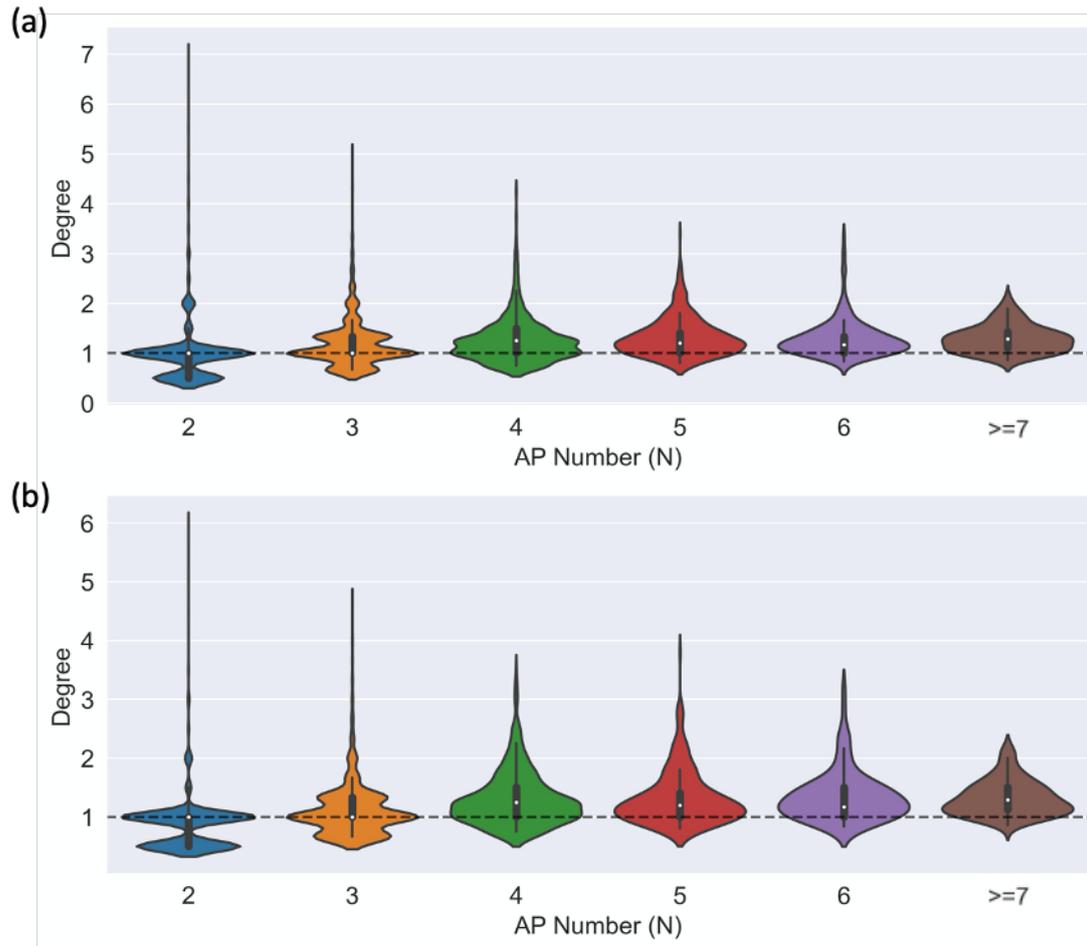

Figure 10. Violin plots for degree of trip chains by node number in (a) Jeonju, (b) Gangneung. The horizontal dash lines correspond to degree value 1.

We group trip chains by different node number. Since the sample size is very small when node >= 7 in both Jeonju (117 individual daily trip chains) and Gangneung (73 individual daily trip chains), we treat samples with node >= 7 as one group in each of the two cities. Figure 10 (a) and (b) generally show a negative correlation between the maximum value of degree and AP number $N$, whereas a slight positive correlation between the minimum value of degree and $N$ is demonstrated. The distribution of degree tends to be more concentrated to the median when the $N$ increases. Moreover, for chain groups with $N=2$ and $N=3$, the medians of degree are 1 in two cities; for chain groups with more $N$, the medians of degree are similar and just over 1. Specially, we can see that for groups with $N=2$ and $N=3$ in two cities, the frequency distributions of degree are similar but rather uneven, which can be attributed to the limitations of $N$. When $N$ is small, there are not many possible topological structures of corresponding trip chains.

The results above can be interpreted from two perspectives. First, for groups with fewer APs, the heterogeneity is more significant due to the more discrete distribution of chain degree. Thus, groups with fewer APs may consist of a larger variety of tourists. On the contrary, groups with more APs appear to be more homogeneous due to the lower extreme high degree value and higher extreme low degree value. This indicates even samples with extreme high degree value are more likely to conduct daily travel with lower degree trip chains. Second, the median values of degree do not significantly increase as the $N$ increases, which implies for groups with more APs, most of them have almost the same travel efficiency as groups with fewer APs. Combining these two perspectives, we discover the principle of least effort and its impacts on tourist travel behavior. The effect is more obvious when AP number $N$ becomes larger. This illustrates even though visitors plan their itinerary with special proclivities, such as unpopular or distant spots, they still tend to choose the most convenient trip chains to achieve the highest travel efficiency.

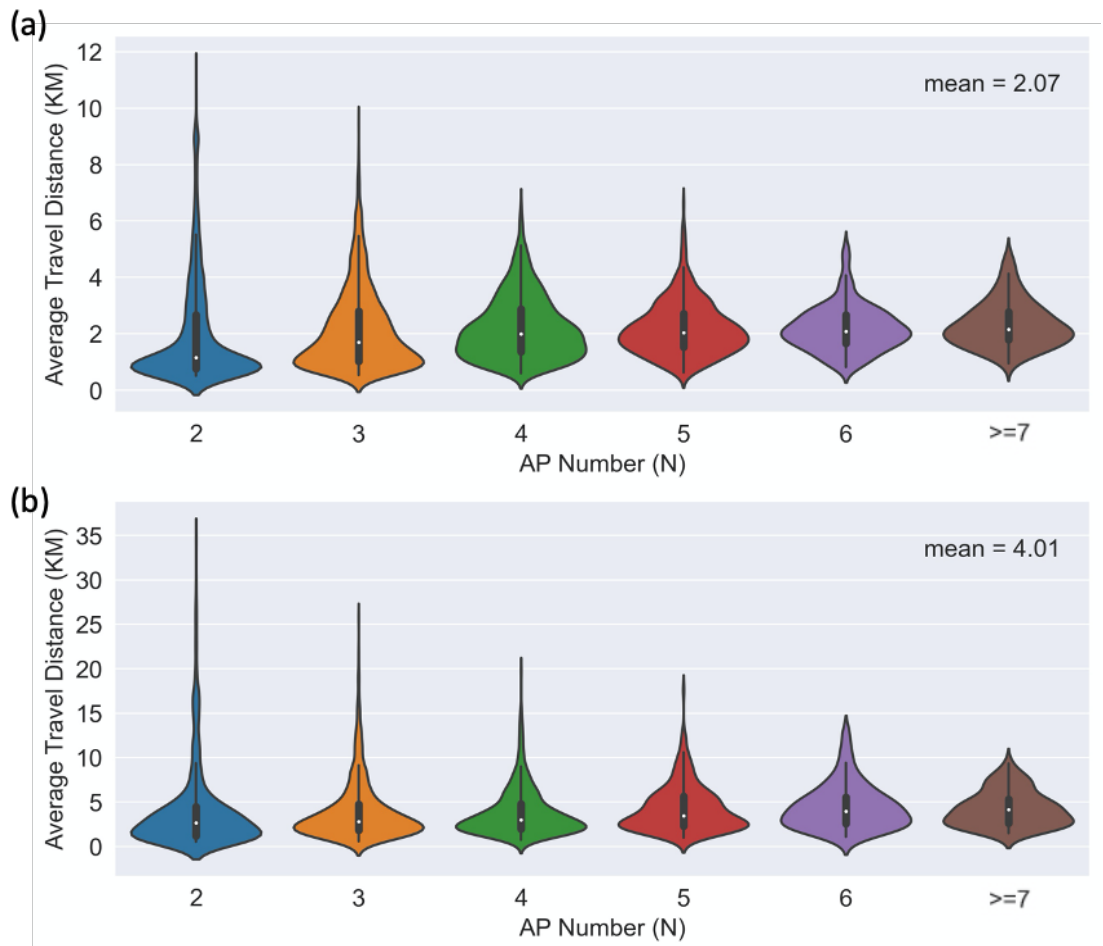

Figure 11. Violin plots for average travel distance of trip chains in (a) Jeonju, (b) Gangneung.

It's worth mentioning that the area of Jeonju and Gangneung is 206.22 km² and 1040 km² respectively. Thus, it makes sense that the mean value of average travel distances of Gangneung (4.01KM) are longer than that of Jeonju (2.07KM). However, such a slight difference implies even though tourist's average distances are relevant to city's area, the activity space of tourists is rather limited. As can be seen from Figure 11, for each city, the maximum value of average travel distance drops as the $N$ increases while the minimum value of average travel distance increases as the $N$ increases. When $N$ gets higher, the distribution of average distance for each group becomes more concentrated. This result also demonstrates the heterogeneity for groups with fewer APs and homogeneity for groups with more APs. Another finding from Figure 11 is that the median values of average travel distance have very slight growth when the $N$ becomes higher for two cities. This indicates although some tourists prefer have trajectories with multiple stops, they will also make their trips as short as impossible, which demonstrates the least effort principle in tourist travel behavior from the perspective of travel distance.

# 6 Conclusions and Implications

Tourism studies suggest that the assumption that travel decisions involving a single destination is misleading. Travelers tend to visit multiple destinations/attractions to meeting their variety-seeking motivations and maximizing benefits from their trips. Hence, the identification of travel patterns involving a range of sites visited has been a critical issue in tourism research (Tideswell & Faulkner, 1999). Along with the transportation literature, this paper proposes tourist daily trip chains to discover underlying travel movement patterns. This model has important theoretical implications for the tourism knowledge base. In fact, there have been several tourism scholars who have mentioned the notion of trip chains, but the attempt to quantitatively identify the structures of trip chains is paucity (e.g., Lau & McKercher, 2006; Stewart & Vogt, 1996). This challenge is somehow attributable to restrictions on accessing proper data enabling researchers to detect individuals' movement in a comprehensive and detailed manner. Most previous studies relied primarily on survey data. This approach, however, contains several limitations (e.g., substantial cost and effort, the potential for response errors; cf. Shoval & Ahas, 2016)

Taking advantage of mobile sensor data providing fine-grained spatiotemporal resolution of travel behaviors, this study discovered 13 key trip chains that account for approximately 76% of "hybrid" and 90% of "intra-city" chain types. The benefits of mobile technology facilitating the collection of digital footprints—within as well as outside a focal destination—makes it possible to explore two types of trip chains including "hybrid" (containing inter-city and intra-city components) and "intra-city" (including only intra-city patterns). The results revealed different formations of trip chains including semantic places and directional movements between two types of trip chains, which demonstrate the complex daily travel network from tourism big data.

Specifically, travelers who show "hybrid" trip chains are likely to prefer to go directly to the hotel after arriving at the focal city and take a rest the first day, and most tourists who pass by the given cities only visit one spot. Based on the result of over 90% measured "intra-city" trip chains, it can be concluded that tourists tend to prefer integrated attractions rather than decentralized and monotonous ones. These findings fill the theoretical gap in tourism literature on multi-destination trips by discovering significant and underlying patterns based upon a full travel trajectory.

Furthermore, this research applies the principle of least efforts (PLE) proposed in evolutionary biology and information systems into travel mobility (Cao et al., 2019). Indeed, the broad theory of PLE serves to explain empirical tourist mobility patterns and regularities exhibited by international travelers. More specifically, this research suggests two indicators—trip chain degree and average travel distance—that quantify individual travel efficiency and reveal the presence of PLE. This framework helps to reveal that travelers make their trips as short as possible and that this affects their structures of trip chains. Tourism researchers have discussed gravity theory (e.g., Morley, Rosselló, & Santana-Gallego, 2014) and distance decay (Lee, Guillet, Law, & Leung, 2012) as means of understanding travel movement behaviors. Importantly, this paper suggests an additional theory to interpret human behaviors (achieving tasks in PLE): travel distance (efforts spent in PLE). It also contributes to methodological approaches to examining these phenomena. Introducing such an analytical framework offers a path toward greater understanding of tourist mobility patterns. It also provides valuable input for many applications (e.g., personalized location-based services for tourism; smart city and smart tourism; and sustainable city planning). In addition, as demonstrated by our analysis results, a constantly growing number of mobile phone data sources contribute a great deal to geographic data mining and knowledge discovery in the age of instant access.

This paper also suggests important managerial implications. The findings discovering underlying trip chains should be beneficial for travel organizers in developing new products. Based on flow-based destination planning (Park, Xu, Jiang, Chen, & Huang,

2020), the structure of trip chains considering directions and sequences of travel movement can become fundamental knowledge in transportation and crowd management as well as the development of travel packages and routes. Recently, DMOs are likely to cooperate with big data firms (e.g., telecommunication companies). This collaboration and cooperation can generate innovative opportunity to access real-time information of travel behaviors, and to collect "big data." Approach to analyzing the trip chain this study suggests should guide for DMOs to not only how to analyze mobile big data, but to better understand travel spatial behaviors, which should be essential to accomplish smart tourism destination.

However, there still exist a few limitations we wish to point out. First, while these documented locations of stop points have higher accuracy than that of years ago, they still contain errors that might mislead the understanding of underlying mobility patterns. Second, our work just examined two cities as study areas. In the future, more cities—especially different types of cities —can be added into our research to achieve generalized results. Another promising direction for future work is that we can infer spatiotemporal information regarding tourist itineraries by mapping the trip chains to the spatial context. In addition, for metropolises with tourists of high average observation days, it is suggested to dig deep into the transition patterns in continuous days.

## Acknowledgements

Thanks for Prof. Yang Xu (Department of Land Survey and Geo-informatics Information, the Hong Kong Polytechnic University), Prof. Sangwon Park (School of Hotel and Tourism Management, the Hong Kong Polytechnic University), and Prof. Wei Tu (School of Architecture and Urban Planning, Shenzhen University) for their advice and revision on this paper.